\documentclass[final]{IEEEtran}
\usepackage{cite}
\usepackage{threeparttable}
\usepackage{graphicx}
\usepackage{picinpar}
\usepackage[cmex10]{amsmath}
\usepackage{amsmath,amsfonts,amssymb}
\usepackage{subfigure}

\usepackage{algorithm}
\usepackage{algorithmic}
\usepackage{stfloats}
\usepackage{bm}

\begin{document}
\newtheorem{lemma}{Lemma}
\newtheorem{corol}{Corollary}
\newtheorem{theorem}{Theorem}
\newtheorem{proposition}{Proposition}
\newtheorem{definition}{Definition}
\newcommand{\e}{\begin{equation}}
\newcommand{\ee}{\end{equation}}
\newcommand{\eqn}{\begin{eqnarray}}
\newcommand{\eeqn}{\end{eqnarray}}
\title{Super-Resolution Sparse MIMO-OFDM Channel Estimation Based on Spatial and Temporal Correlations}

\author{Zhen Gao, Linglong Dai,~\IEEEmembership{Member,~IEEE}, Zhaohua Lu,\\ Chau Yuen,~\IEEEmembership{Senior Member,~IEEE},
and Zhaocheng Wang,~\IEEEmembership{Senior Member,~IEEE}%
 \vspace*{-3.0mm}
\thanks{Manuscript received on 22 January 2014.}
 \thanks{Z. Gao, L. Dai, and Z. Wang are with Tsinghua National Laboratory for
 Information Science and Technology (TNList), Department of Electronic Engineering,
 Tsinghua University, Beijing 100084, China (E-mails: gao-z11@mails.tsinghua.edu.cn; \{daill, zcwang\}@mail.tsinghua.edu.cn).}
 \thanks{Z. Lu is with ZTE Corporation, Shenzhen 518000, China (E-mail: lu.zhaohua@zte.com.cn).}  %
 \thanks{C. Yuen is with Singapore University of Technology and Design,
Singapore 138682 (E-mail: yuenchau@sutd.edu.sg).}  %
\thanks{This work was supported by National Key Basic Research Program of
China (Grant No. 2013CB329203), National Nature Science Foundation
of China (Grant No. 61271266), National High Technology Research
and Development Program of China (Grant No. 2014AA01A704), and the ZTE fund project (Grant No. CON1307250001).}  %
}

\maketitle
\begin{abstract}
This letter proposes a parametric sparse multiple input multiple output (MIMO)-OFDM channel estimation scheme based on the finite rate of innovation (FRI) theory, whereby super-resolution estimates of path delays with arbitrary values can be achieved. Meanwhile, both the spatial and temporal correlations of wireless MIMO channels are exploited to improve the accuracy of the channel estimation. For outdoor communication scenarios, where wireless channels are sparse in nature, path delays of different transmit-receive antenna pairs share a common sparse pattern due to the spatial correlation of MIMO channels. Meanwhile, the channel sparse pattern is nearly unchanged during several adjacent OFDM symbols due to the temporal correlation of MIMO channels. By simultaneously exploiting those MIMO channel characteristics, the proposed scheme performs better than existing state-of-the-art schemes. Furthermore, by joint processing of signals associated with different antennas, the pilot overhead can be reduced under the framework of the FRI theory.
\end{abstract}
 \vspace*{-1mm}
\begin{IEEEkeywords}
Super-resolution, sparse channel estimation, MIMO-OFDM, finite rate of innovation (FRI), spatial and temporal correlations.
\end{IEEEkeywords}
 \vspace*{-2.0mm}
\IEEEpeerreviewmaketitle
\section{Introduction}
\IEEEPARstart{M}{ULTIPLE} input multiple output (MIMO)-OFDM is widely recognized as a key technology for future wireless communications due to its high spectral efficiency and superior robustness to multipath fading channels \cite{{4g}}.

For MIMO-OFDM systems, accurate channel estimation is essential to guarantee the system performance \cite{FP}. Generally, there are two categories of channel estimation scheme for MIMO-OFDM systems. The first one is nonparametric scheme, which adopts orthogonal frequency-domain pilots or orthogonal time-domain training sequences to convert the channel estimation in MIMO systems to that in single antenna systems \cite{{FP}}. However, such scheme suffers from high pilot overhead when the number of transmit antennas increases. The second category is parametric channel estimation scheme, which exploits the sparsity of wireless channels to reduce the pilot overhead \cite{{CS_Sparse},{MIMOOFDM}}. The parametric scheme is more favorable for future wireless systems as it can achieve higher spectral efficiency. However, path delays of sparse channels are assumed to be located at the integer times of the sampling period, which is usually unrealistic in practice.

In this letter, a more practical sparse MIMO-OFDM channel estimation scheme based on spatial and temporal correlations of sparse wireless MIMO channels is proposed to deal with arbitrary path delays. The main contributions of this letter are summarized as follows.
First, the proposed scheme can achieve super-resolution estimates of arbitrary path delays, which is more suitable for wireless channels in practice.
Second, due to the small scale of the transmit and receive antenna arrays compared to the long signal transmission distance in typical MIMO antenna geometry, channel impulse responses (CIRs) of different transmit-receive antenna pairs share common path delays \cite{{SCS}}, which can be translated as a common sparse pattern of CIRs due to the spatial correlation of MIMO channels. Meanwhile, such common sparse pattern is nearly unchanged along several adjacent OFDM symbols due to the temporal correlation of wireless channels \cite{{IT},{CS_TSP}}. Compared with previous work which just simply extends the sparse channel estimation scheme in single antenna systems to that in MIMO by exploiting the spatial correlation of MIMO channels \cite{{SCS}} or only considers the temporal correlation for single antenna systems \cite{{IT},{CS_TSP}}, the proposed scheme exploits both spatial and temporal correlations to improve the channel estimation accuracy.
Third, we reduce the pilot overhead by using the finite rate of innovation (FRI) theory \cite{FRI}, which can recovery the analog sparse signal with very low sampling rate, as a result, the average pilot overhead per antenna only depends on the channel sparsity level instead of the channel length.


\textit{Notation}: $(\cdot)^\dag $ and $(\cdot)^H $ are the Moore-Penrose matrix inversion operation and matrix conjugate transpose operation, respectively. ${\rm{diag\{\bf{x}}\} }$ is a diagonal matrix with the vector $\bf{x}$ on its diagonal. The operator $*$ denotes the linear convolution.
\section{Sparse MIMO Channel Model}
\begin{figure}
     \centering
     \includegraphics[width=8.3cm, keepaspectratio]%
     {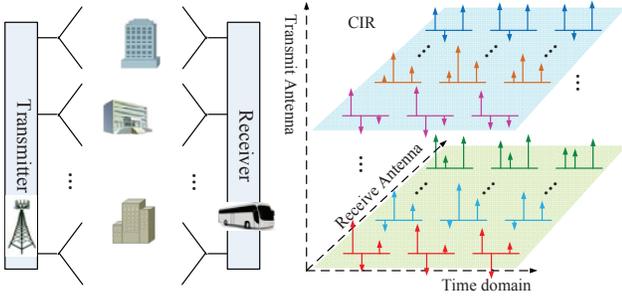}
     \vspace*{-1.5mm}
     \caption{Spatial and temporal correlations of wireless MIMO channels. 
     }
     \label{fig:str}
     \vspace*{-4mm}
\end{figure}
The MIMO channel is shown in Fig. \ref{fig:str}, and its following characteristics will be considered in this letter.
\subsubsection{Channel Sparsity}
In typical outdoor communication scenarios, the CIR is intrinsically sparse due to several significant scatterers \cite{{SCS},{CS_Sparse}}. For an ${N_t}\times  {N_r}$ MIMO system, the CIR ${h^{(i,j)}}(t)$ between the $i$th transmit antenna and the $j$th receive antenna can be modelled as \cite{4g},
\begin{equation}
\begin{small}
\begin{array}{l}
\!\!\!\!\!{h^{(i,j)}}(t) = \sum\limits_{p = 1}^P {\alpha _{p}^{(i,j)}\delta (t  - \tau _{p}^{(i,j)})},{\kern 2pt}1 \le i \le {N_t},{\kern 2pt}1 \le j \le {N_r},\label{equ:CIR}
\end{array}
\end{small}
\end{equation}
where $\delta (\cdot )$ is Dirac function, $P$ is the total number of resolvable propagation paths, $\tau _{p}^{(i,j)}$ and ${\alpha _{p}^{(i,j)}}$ denote the path delay and path gain of the $p$th path, respectively.
\subsubsection{Spatial Correlation}
Because the scale of the transmit or receive antenna array is very small compared to the long signal transmission distance, channels of different transmit-receive antenna pairs share very similar scatterers. Meanwhile, for most communication systems, the path delay difference from the similar scatterer is far less than the system sampling period. Therefore, CIRs of different transmit-receive antenna pairs share a common sparse pattern, although the corresponding path gains may be quite different\cite{SCS}.
\subsubsection{Temporal Correlation}
For wireless channels, the path delays vary much slowly than the path gains, and the path gains vary continuously \cite{IT}. Thus, the channel sparse pattern is nearly unchanged during several adjacent OFDM symbols, and the path gains are also correlated \cite{CS_TSP}.
\section{Sparse MIMO-OFDM Channel Estimation}%
In this section, the widely used pilot pattern is briefly introduced at first, based on which a super-resolution sparse MIMO-OFDM channel estimation method is then applied. Finally, the required number of pilots is discussed under the framework of the FRI theory.
\vspace*{-2mm}
\subsection{Pilot Pattern}
\begin{figure}[!tp]
     \centering
     \includegraphics[width=2.5cm, keepaspectratio,angle=90]
     {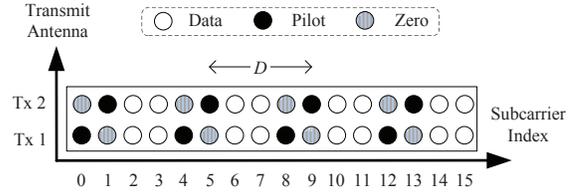}
     \vspace*{-2mm}
    \caption{Pilot pattern. Note that the specific $N_t=2$, $D=4$, $N_p=4$, $N_{p\_\text{total}}=8$ are used for illustration purpose.}
     \label{fig:pilot}
    \vspace*{-4mm}
\end{figure}

The pilot pattern widely used in common MIMO-OFDM systems is illustrated in Fig. \ref{fig:pilot}. In the frequency domain, $N_p$ pilots are uniformly spaced with the pilot interval $D$ (e.g., $D=4$ in Fig. \ref{fig:pilot}). Meanwhile, every pilot is allocated with a pilot index $l$ for $0 \le l \le N_p-1 $, which is ascending with the increase of the subcarrier index.
Furthermore, to distinguish MIMO channels associated with different transmit antennas, each transmit antenna uses a unique subcarrier index initial phase ${\theta _i}$ for $1 \le i \le {N_t}$ and $(N_t-1)N_p$ zero subcarriers to ensure the orthogonality of pilots \cite{MIMOOFDM}. Therefore, for the $i$th transmit antenna, the subcarrier index of the $l$th  pilot is
\begin{align}
I_{{\rm{pilot}}}^i(l) = {\theta _i} + lD,{\kern 3pt}0 \le l \le N_p-1.\label{equ:pilot_index}
\end{align}
Consequently, the total pilot overhead per transmit antenna is~$N_{p\_\text{total}}=N_tN_p$, and thus $N_p$ can be also referred as the average pilot overhead per transmit antenna in this letter.
\subsection{Super-Resolution Channel Estimation}\label{2c}
At the receiver, the equivalent baseband channel frequency response (CFR) $H(f)$ can be expressed as
\begin{align}
\begin{array}{l}
H(f) =\sum\limits_{p = 1}^P {{\alpha _p}{e^{ - j2\pi f{\tau _p}}}}, {\kern 3pt} -f_s/2 \le f \le f_s/2,
\end{array}
\label{equ:freq}
\end{align}
where superscripts $i$ and $j$ in (\ref{equ:CIR}) are omitted for convenience,~$f_s=1/T_s$ is the system bandwidth, and $T_s$ is the sampling period. Meanwhile, the $N$-point discrete Fourier transform (DFT) of the time-domain equivalent baseband channel can be expressed as \cite{SCS}, i.e.,
\begin{align}
\begin{small}
H[k]=H(\frac{{k{f_s}}}{N}),{\kern 3pt}0 \le k \le N-1.\label{equ:DFT}
\end{small}
\end{align}
Therefore, for the $(i,j)$th transmit-receive antenna pair, according to (\ref{equ:pilot_index}) - (\ref{equ:DFT}), the estimated CFRs over pilots can be written as
\begin{align}
\begin{small}
\begin{array}{l}
{\hat {\cal H}}^{(i,j)}[l]= H[I_{{\rm{pilot}}}^i(l)] = H(\frac{{({\theta _i} + lD){f_s}}}{N})\\
 = \sum\limits_{p = 1}^P {\alpha _p^{(i,j)}} {e^{ - j2\pi \frac{{({\theta _i} + lD){f_s}}}{N}\tau _p^{(i,j)}}} + {W^{(i,j)}}[l],
\end{array}\label{equ:CIR_m+1}
\end{small}
\end{align}
where ${{\hat {\cal H}}^{(i,j)}}[l]$ for $0 \le l \le N_p-1$ can be obtained by using the conventional minimum mean square error (MMSE) or least square (LS) method \cite{FP}, and ${W^{(i,j)}[l]}$ is the additive white Gaussian noise (AWGN).

Eq. (\ref{equ:CIR_m+1}) can be also written in a vector form as
\begin{align}
\begin{small}
{\hat {\cal H}}^{(i,j)}[l] = ({{\bf{v }}^{(i,j)}}[l])^{T}{\bf{a }}^{(i,j)}+W^{(i,j)}[l],\label{equ:CIR_m+2}
\end{small}
\end{align}
where ${\bf{v}}^{(i,j)}[l] = {[{\gamma ^{lD\tau _{1}^{(i,j)}}},{\gamma ^{lD\tau _{2}^{(i,j)}}}, \cdots ,{\gamma ^{lD\tau _{P}^{(i,j)}}}]^T}$, ${\bf{a}}^{(i,j)} = {[\alpha _{1}^{(i,j)}{\gamma ^{{\theta _i}\tau _{1}^{(i,j)}}},\alpha _{2}^{(i,j)}{\gamma ^{{\theta _i}\tau _{2}^{(i,j)}}}, \cdots ,\alpha _{P}^{(i,j)}{\gamma ^{{\theta _i}\tau _{P}^{(i,j)}}}]^T}$, and $\gamma  = {e^{ - j2\pi \frac{{{f_s}}}{N}}}$.

Because the wireless channel is inherently sparse and the small scale of multiple transmit or receive antennas is negligible compared to the long signal transmission distance, CIRs of different transmit-receive antenna pairs share common path delays, which is equivalently translated as a common sparse pattern of CIRs due to the spatial correlation of MIMO channels \cite{SCS}, i.e., $\tau _{p}^{(i,j)} = \tau _{p}$ and ${\bf{v }}^{(i,j)}[l]={\bf{v}}[l]$ for~$1 \le p \le P$, $1 \le i \le {N_t}$, $1 \le j \le {N_r}$. Hence by exploiting such spatially common sparse pattern shared among different receive antennas associated with the $i$th transmit antenna, we have
\begin{align}
  \vspace*{-2mm}
{{\bf{\hat H}}^i} = {{\bf{V}}}{{\bf{A}}^i} + {{\bf{W}}^i},{\kern 6pt}1 \le i \le {N_t},\label{equ:super1}
\end{align}
where the $N_p\times N_r$ measurement matrix ${{\bf{\hat H}}^i}$ is
\begin{equation}
\footnotesize
\!\!{{\bf{\hat H}}^i} = \left[ {\begin{array}{*{20}{c}}
{\hat{\cal H}^{(i,1)}[0]}&{\hat{\cal H}^{(i,2)}[0]}& \cdots &{\hat{\cal H}^{(i,{N_r})}[0]}\\
{\hat{\cal H}^{(i,1)}[1]}&{\hat{\cal H}^{(i,2)}[1]}& \cdots &{\hat{\cal H}^{(i,{N_r})}[1]}\\
 \vdots & \vdots & \ddots & \vdots \\
{\hat{\cal H}^{(i,1)}[{N_p} - 1]}&{\hat{\cal H}^{(i,2)}[{N_p} - 1]}& \cdots &{\hat{\cal H}^{(i,{N_r})}[{N_p} - 1]}
\end{array}} \right],\nonumber
\end{equation}
${\bf{V}}={[{\bf{v}}^{}[0],{\bf{v}}^{}[1],\cdots,{\bf{v}}^{}[{N_p}-1]]^T}$ is a Vandermonde matrix of size $N_p \times N_p$, ${\bf{A}}^i=[{\bf{a}}^{(i,1)},{\bf{a}}^{(i,2)},\cdots,{\bf{a}}^{(i,{N_r})}]$ of size $N_p \times N_r$, and ${\bf{W}}^{i}$ is an $N_p\times N_r$ matrix with ${W^{(i,j)}[l]}$ in its $j$th column and the $(l+1)$th row.


When all $N_t$ transmit antennas are considered based on (\ref{equ:super1}), we have
\begin{align}
\begin{small}
{{\bf{\hat H}}} = {{\bf{V}}}{{\bf{A}}} + {{\bf{W}}},\label{equ:super2}
\end{small}
\end{align}
where ${{{\bf{\hat H}}}} = [{\bf{\hat H}}^1,{\bf{\hat H}}^2, \cdots ,{\bf{\hat H}}^{{N_t}}]$ of size $N_p\times N_tN_r$, ${{{\bf{ A}}}} = [{\bf{ A}}^1,{\bf{ A}}^2, \cdots ,{\bf{ A}}^{N_t}]$, and ${{{\bf{ W}}}} = [{\bf{ W}}^1,{\bf{ W}}^2, \cdots ,{\bf{ W}}^{{N_t}}]$.

Comparing the formulated problem (\ref{equ:super2}) with the classical direction-of-arrival (DOA) problem \cite{esprit}, we find out that they are mathematically equivalent. Specifically, the traditional DOA problem is to typically estimate the DOAs of the $P$ sources from a set of time-domain measurements, which are obtained from the $N_p$ sensors outputs at $N_tN_r$ distinct time instants (time-domain samples). In contrast to our problem in~(\ref{equ:super2}), we try to estimate the path delays of $P$ multipaths from a set of frequency-domain measurements, which are acquired from $N_p$ pilots of $N_tN_r$ distinct antenna pairs (antenna-domain samples). 
It has been verified in \cite{Eldar} that the total least square estimating signal parameters via rotational invariance techniques (TLS-ESPRIT) algorithm in~\cite{esprit} can be applied to~(\ref{equ:super2}) to efficiently estimate path delays with arbitrary values.

By using the TLS-ESPRIT algorithm, we can obtain super-resolution estimates of path delays, i.e., $\hat {\tau} _{p}$ for $1 \le p \le P$, and thus ${\bf{\hat V }}$ can be obtained accordingly. Then path gains can be acquired by the LS method \cite{CS_TSP}, i.e.,
\begin{align}
{\bf{\hat A}} = {\bf{\hat V }}^{ \dag }{\bf{\hat H}} = {({\bf{\hat V}}^{H}{\bf{\hat V }})^{ - 1}}{\bf{\hat V }}^{H}{\bf{\hat H}}.
\end{align}
For a certain entry of ${\bf{\hat A}}$, i.e., ${{\hat \alpha} _{p}^{(i,j)}{\gamma ^{{\theta _i}{\hat \tau} _{p}}}}$, because ${\theta _i}$ is known at the receiver and ${\hat \tau} _{p}$ has been estimated after applying the TLS-ESPRIT algorithm, we can easily obtain the estimation of the path gain ${\hat \alpha} _{p}^{(i,j)}$ for $1 \le p \le P$, $1 \le i \le {N_t}$, $1 \le j \le {N_r}$. Finally, the complete CFR estimation over all OFDM subcarriers can be obtained based on (\ref{equ:freq}) and (\ref{equ:DFT}).

Furthermore, we can also exploit the temporal correlation of wireless channels to improve the accuracy of the channel estimation. First, path delays of CIRs during several adjacent OFDM symbols are nearly unchanged \cite{{IT},{CS_TSP}}, which is equivalently referred as a common sparse pattern of CIRs due to the temporal correlation of MIMO channels. Thus, the Vandermonde matrix $\bf{V}$ in (\ref{equ:super2}) remains unchanged across several adjacent OFDM symbols. Moreover, path gains during adjacent OFDM symbols are also correlated owing to the temporal continuity of the CIR, so ${{\bf{A}}}$'s in (\ref{equ:super2}) for several adjacent OFDM symbols are also correlated. Therefore, when estimating CIRs of the $q$th OFDM symbol, we can jointly exploit ${\hat{\bf{H}}}$'s of several adjacent OFDM symbols based on (\ref{equ:super2}), i.e.,
\begin{align}
\begin{small}
\frac{{\sum\limits_{\rho = q - R}^{q + R} {{{{\bf{\hat H}}}_{\rho}}} }}{{2R + 1}} = {{\bf{V}}_{q}}\frac{{\sum\limits_{{\rho} = q - R}^{q + R} {{{\bf{A}}_{\rho}}} }}{{2R + 1}} + \frac{{\sum\limits_{{\rho} = q - R}^{q + R} {{{\bf{W}}_{\rho}}} }}{{2R + 1}}. \label{equ:esprit}
\end{small}
\end{align}
where the subscript $\rho$ is used to denote the index of the OFDM symbol, and the common sparse pattern of CIRs is assumed in~$2R+1$ adjacent OFDM symbols \cite{CS_TSP}. In this way, the effective noise can be reduced, so the improved channel estimation accuracy is expected.

In contrast to the existing nonparametric scheme which estimates the channel by interpolating or predicting based on CFRs over pilots \cite{{4g},{FP}}, our proposed scheme exploits the sparsity as well as the spatial and temporal correlations of wireless MIMO channels to first acquire estimations of channel parameters including path delays and gains, and then obtain the estimation of CFR according to  (\ref{equ:freq}) and (\ref{equ:DFT}). 
\subsection{Discussion on Pilot Overhead}
Compared with the model of the multiple filters bank based on the FRI theory \cite{Eldar}, it can be found out that CIRs of $N_tN_r$ transmit-receive antenna pairs are equivalent to the $N_tN_r$ semi-period sparse subspaces, and the $N_p$ pilots are equivalent to the $N_p$ multichannel filters. Therefore, by using the FRI theory, the smallest required number of pilots for each transmit antenna is $N_p=2P$ in a noiseless scenario. For practical channels with the maximum delay spread ${\tau _{\max }}$, although the normalized channel length $L={\tau _{\max }}/T_s$ is usually very large, the sparsity level $P $ is small, i.e., $P \ll L$ \cite{CS_Sparse}. Consequently, in contrast to the nonparametric channel estimation method where the required number of pilots heavily depends on $L$, our proposed parametric scheme only needs $2P$ pilots in theory. Note that the number of pilots in practice is larger than $2P$ to improve the accuracy of the channel estimation due to AWGN.

\section{Simulation Results}
A simulation study was carried out to compare the performance of the proposed scheme with those of the existing state-of-the-art methods for MIMO-OFDM systems. The conventional comb-type pilot and time-domain training based orthogonal pilot (TTOP) \cite{FP} schemes were selected as the typical examples of the nonparametric channel estimation scheme, while the recent time-frequency joint (TFJ) channel estimation scheme \cite{MIMOOFDM} was selected as an example of the conventional parametric scheme. System parameters were set as follows: the carrier frequency is $f_c=1{\kern 2pt}\text{GHz}$, the system bandwidth is~$f_s=10 {\kern 2pt}\text{MHz}$, the size of the OFDM symbol is $N=4096$, and $N_g=256$ is the guard interval length, which can combat channels whose maximum delay spread is $25.6{\kern 2pt}\mu { s}$. The International Telecommunication Union Vehicular B (ITU-VB) channel model with the maximum delay spread $20{\kern 2pt}\mu { s}$ and the number of paths $P=6$ \cite{MIMOOFDM} was considered.

\begin{figure}[!tp]
     \centering
     \includegraphics[width=7cm, keepaspectratio]
     {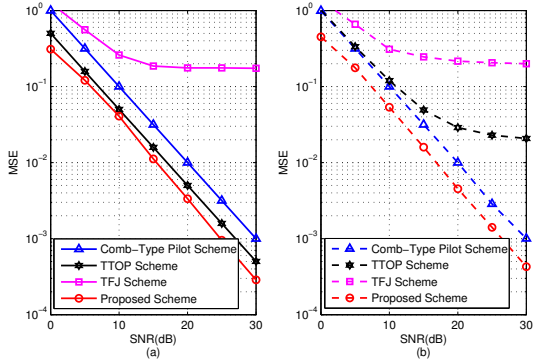}
           \vspace*{-1.2mm}
    \caption{MSE performance comparison of different schemes in a $4\times4$ MIMO system. (a) Static channel; (b) Time-varying channel with the mobile speed of 90 km/h.}
     \label{fig:MSE1}
           \vspace*{-1.3mm}
\end{figure}

Fig. \ref{fig:MSE1} compares the mean square error (MSE) performance of different channel estimation schemes. Both the static ITU-VB channel and time-varying ITU-VB channel with the mobile speed of 90 km/h in a $4\times4$ MIMO system were considered. The comb-type pilot based scheme used  $N_p=256$ pilots, the TTOP scheme used $N_p=64$ pilots with $T$ adjacent OFDM symbols for training where $T=4$ for the time-varying channel and $T=8$ for the static channel to achieve better performance, the TFJ scheme adopted time-domain training sequences of 256-length and $N_p=64$ pilots, and our proposed scheme used $N_p=64$ pilots with $R=4$ for fair comparison. From Fig. \ref{fig:MSE1}, we can observe that the conventional parametric TFJ scheme is inferior to other three schemes obviously. Meanwhile, for static ITU-VB channel, the MSE performance of the proposed parametric scheme is more than 2 dB and 5 dB better than the TTOP and comb-type pilot based schemes, respectively. Moreover, for the time-varying ITU-VB channel, the superior performance of our proposed parametric scheme to conventional nonparametric schemes is more obvious. The existing sparse channel estimation scheme \cite{MIMOOFDM} does not work, because path delays may not be located at the integer times of the sampling period for practical channels. The TTOP scheme works well over static channels, but it performs poorly over fast time-varying channels, since it assumes the channel is static during the adjacent OFDM symbols. Finally, the comb-type pilot based scheme performs worse than our proposed scheme, and it also suffers from much higher pilot overhead.

\begin{figure}[!tp]
     \centering
     \includegraphics[width=6cm, keepaspectratio]
     {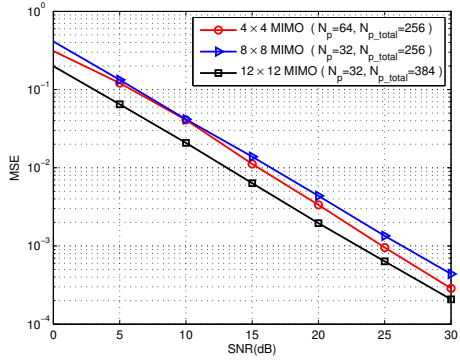}
     \vspace*{-1.2mm}
    \caption{MSE performance of the proposed scheme in $4\times4$, $8\times8$ and $12\times12$ MIMO systems.}
     \label{fig:MSE2}
      \vspace*{-4.5mm}
\end{figure}

Fig. \ref{fig:MSE2} compares the MSE performance of the proposed scheme in $4\times4$, $8\times8$ and $12\times12$ MIMO systems. We can observe that the MSE performance of the proposed scheme in~$12\times12$ MIMO system is superior to that in $8\times8$ MIMO system by 5 dB with the same $N_p$, and outperforms that in~$4\times4$ MIMO system with the reduced $N_p$. These simulations indicate that with the increased number of antennas, the MSE performance improves with the same $N_p$. Equivalently, to achieve the same channel estimation accuracy, the required number of pilots $N_p$ can be reduced. As a result, the total pilot overhead $N_{p\_\text{total}}$ in our proposed scheme does not increase linearly with the number of transmit antennas $N_t$ because the required $N_p$ reduces when $N_t$ increases accordingly. The reason is that with the increased number of antennas, the dimension of the measurement matrix (e.g., ${{\bf{\hat H}}}$ in (\ref{equ:super2})) in the TLS-ESPRIT algorithm or the number of the sampling in the model of multiple filters bank \cite{Eldar} increases, thus the accuracy of the path delay estimate improves accordingly.

The superior performance of the proposed scheme is contributed by following reasons. First, the spatially common sparse pattern shared among CIRs of different transmit-receive antenna pairs is exploited in the proposed scheme, such that we can employ the TLS-ESPRIT algorithm to obtain super-resolution estimations of path delays with arbitrary values. Meanwhile, the FRI theory indicates that the smallest required number of pilots is $N_p=2P$ in a noiseless scenario. Therefore, the pilot overhead can be reduced as compared with conventional nonparametric schemes. Second, our scheme exploits the temporal correlation of wireless channels, namely, across several adjacent OFDM symbols, the sparse pattern of the CIR remains unchanged, and path gains are also correlated. Accordingly, by joint processing of signals of adjacent OFDM symbols based on (\ref{equ:super2}), the effective noise can be reduced and thus the accuracy of the channel estimation is improved further.
\vspace*{-1mm}
\section{Conclusions}
The proposed super-resolution sparse MIMO channel estimation scheme exploits the sparsity as well as the spatial and temporal correlations of wireless MIMO channels. It can achieve super-resolution estimates of path delays with arbitrary values, and has higher channel estimation accuracy than conventional schemes. Under the framework of the FRI theory, the required number of pilots in the proposed scheme is obviously less than that in nonparametric channel estimation schemes. Moreover, simulations demonstrate that the average pilot overhead per transmit antenna will be interestingly reduced with the increased number of antennas.


\begin{thebibliography}{10}
\bibitem{4g}G. Stuber, J. Barry, S. Mclaughlin, Y. Li, M. Ingram, and T. Pratt,
``Broadband MIMO-OFDM wireless communications,"
{\em Proc. IEEE,} vol. 92, no. 2, pp. 271-294, Feb. 2004.







\bibitem{FP}I. Barhumi, G. Leus, and M. Moonen,
``Optimal training design for MIMO OFDM systems in mobile wireless channels,"
{\em IEEE Trans. Signal Process.}, vol. 3, no. 6, pp. 958-974, Dec. 2009.




\bibitem{CS_Sparse}W. U. Bajwa, J. Haupt, A. M. Sayeed, and R. Nowak,
``Compressed channel sensing: A new approach to estimating sparse multipath channels,"
{\em Proc. IEEE}, vol. 98, no. 6, pp. 1058-1076, Jun. 2010.

\bibitem{MIMOOFDM}L. Dai, Z. Wang, and Z. Yang,
``Spectrally efficient time-frequency training OFDM for mobile large-scale MIMO systems,"
{\em IEEE J. Sel. Areas Commun.}, vol. 31, no. 2, pp. 251-263, Feb. 2013.

\bibitem{SCS}Y. Barbotin and M. Vetterli,
``Estimation of sparse MIMO channels with common support,"
{\em IEEE Trans. Commun.}, vol. 60, no. 12, pp. 3705-3716, Dec. 2012.



\bibitem{IT}I. Telatar and D. Tse,
``Capacity and mutual information of wideband multipath fading channels,"
{\em IEEE Trans. Inf. Theory}, vol. 46, no. 4, pp. 1384-1400, Jul. 2000.


\bibitem{CS_TSP}L. Dai, J. Wang, Z. Wang, P. Tsiaflakis, and M. Moonen,
``Spectrum- and energy-efficient OFDM based on simultaneous multi-channel reconstruction,"
{\em IEEE Trans. Signal Process.}, vol. 61, no. 23, pp. 6047-6059, Dec. 2013.


\bibitem{FRI}P. L. Dragotti, M. Vetterli, and T. Blu,
``Sampling moments and reconstructing signals of finite rate of innovation: Shannon meets Strang-Fix,"
{\em IEEE Trans. Signal Process.}, vol. 55, no. 5, pp. 1741-1757, May. 2007.



\bibitem{esprit}R. Roy and T. Kailath,
``ESPRIT-estimation of signal parameters via rotational invariance techniques,"
{\em IEEE Trans. Acoust., Speech, Signal Process.}, vol. 37, no. 7, pp. 984-995, Jul. 1989.%


\bibitem{Eldar}K. Gedlyahu and Y. C. Eldar,
``Time-delay estimation from low-rate samples: A union of subspaces approach,"
{\em IEEE Trans. Signal Process.}, vol. 58, no. 6, pp. 3017-3031, Sep. 2010.






\end{thebibliography}
\end{document}